\documentclass{JHEP3} 
\title{Space/Time Noncommutativity
in String Theories without Background Electric Field
\thanks{Work supported in part by INFN and MURST of Italy.}}

\author{Giuseppe De Risi\\Dipartimento di Fisica and Sezione 
I.N.F.N., Universit\`a di Perugia, Via A. Pascoli I-06123, 
Perugia, Italia. \email{E-mail:derisi$@$pg.infn.it}}

\author{Gianluca Grignani\\Dipartimento di Fisica and Sezione 
I.N.F.N., Universit\`a di Perugia, Via A. Pascoli I-06123, 
Perugia, Italia. \email{E-mail:grignani$@$pg.infn.it} }

\author{Marta Orselli\\Dipartimento di Fisica and I.N.F.N. Gruppo 
Collegato di Parma, Universit\`a di Parma,
Parco Area delle Scienze 7A I-43100 Parma, 
Italia. \email{E-mail:orselli$@$fis.unipr.it}}

\abstract{The appearance of space/time non-commutativity in theories 
of open strings with a constant non-diagonal background 
metric is considered. 
We show that, even if the space-time 
coordinates commute, when there is a metric with a time-space component,
no electric field and the boundary condition along the spatial 
direction is Dirichlet, 
a Moyal phase still arises in products of vertex operators. 
The theory is in fact dual to the non-commutatitive open string 
(NCOS) theory. The correct definition of the 
vertex operators for this theory is provided.
We study the system also in the presence of a $B$ field. 
We consider the case in which the Dirichlet spatial direction is compactified
and analyze the effect of these background 
on the closed string spectrum. 
We then heat up the system.
We find that the Hagedorn temperature 
depends in a non-extensive way on the parameters of the background and 
it is the same for the closed and the open string sectors.} 


\preprint{UPRF-2002-16} 

\begin{document} 
\def\be{\begin{equation}} 
\def\ee{\end{equation}} 
\def\bea{\begin{eqnarray}} 
\def\eea{\end{eqnarray}} 
\def\nn{\nonumber} 
\def\const{{\rm const}} 
\def\v{\varphi} 
\def\s {\sigma} 
\def\a {\alpha}
\def\t {\tau}
\def\vcl{\varphi_{\rm cl}} 
\newcommand{\no}[1]{:\!#1\!:} 
\def\la{\left\langle} 
\def\ra{\right\rangle} 
\def\d{\partial} 
\def\se{S_{\rm eff}} 
\def\tr{\rm Tr}

\section{Introduction} 

Noncommutativity in open string theory has been 
studied since Witten's seminal paper on open string field theory 
\cite{Witten:1985cc}.
Recently, there has been much progress in
understanding the low energy description of strings and D-branes
in electromagnetic backgrounds~\cite{Abouelsaood:gd,Ambjorn:2000yr,
Douglas:2001ba} and how the string dynamics is described 
by a Yang-Mills field theory with space/space 
noncommutativity~\cite{Seiberg:1999vs}.
When D-branes are placed in a background electric field, 
noncommutativity occurs between time and space coordinates 
\cite{Seiberg:2000ms,Gopakumar:2000na}. There is a critical value of the 
electric field beyond which the theory does not make sense. 
Since the non-commutative scale is intrinsically tied to the string scale,
by approaching the critical field the theory does not become a
non-commutative field theory~\footnote{
Field theories with space/time noncommutativity  
have inconsistencies related to the lack of unitarity, see for example
\cite{Gomis:2000bn,Seiberg:2000gc,Barbon:2000sg,
Alvarez-Gaume:2001ka,Bassetto:2001vf}.}. 
However, it is possible to consider a particular limit in which 
the closed strings, and therefore gravity, decouple, leaving a more 
tractable theory of only open strings. These theories are known as $p+1$ Non 
Commutative Open String Theories (NCOS) \cite{Seiberg:2000ms, 
Gopakumar:2000na}, where $p$ is the dimension of the D$p$-brane.

It was then argued \cite{Klebanov:2000pp} that 
when there is a compactified direction, closed strings do not 
decouple from the spectrum in the NCOS limit. The electric field 
tends to move apart the open string ends but, since the direction 
is compactified, they can join with a finite probability
after encircling completely the compactified direction 
and form again a closed string. This explains also the fact 
that only closed strings with strictly positive winding 
number are allowed, strings can wind only in one sense.
At first sight it may seem strange that the presence of 
the electric field will change the dynamics of closed 
strings which are neutral. This is due to the fact that the electric 
field can be turned into a 
background Neveu-Schwarz $B$ field by gauge transformations. The 
argument in \cite{Klebanov:2000pp} was then extended  
in \cite{Gomis:2000bd} and in \cite{Danielsson:2000gi}, where the NCOS
limit of a type IIA/B superstring theory was considered and a new sector 
of string theory discovered. These are the Wound String theory of 
\cite{Danielsson:2000gi} and the Non Relativistic Closed String 
theory of \cite{Gomis:2000bd}. These are only closed strings 
in the absence of D-branes but when D-branes are present  
represent a generalization of NCOS theories.

In this Paper we shall show that noncommutativity 
in open string theory can arise also
when only a metric with an appropriate form is present 
and one of the spatial direction has Dirichlet boundary conditions.
The theory we consider
is dual to the one studied in~\cite{Seiberg:2000ms,Gopakumar:2000na,
Klebanov:2000pp}. 
In fact, when only a Neveu-Schwarz $B$-field in the directions $B_{0i}$ 
is present, the Buscher rules for duality
\cite{Buscher:1988qj, Buscher:1987sk, Giveon:1994fu}
tell us that the dual theory contains only a nontrivial metric and 
no $B$ field at all. Since there is no
$B$-field the space-time coordinates commute, so how noncommutative effects  
might arise in this situation is a non-trivial question. 
We will show that the source of noncommutativity can be found in the 
Moyal phase which appears in the computation of scattering 
amplitudes of open string vertex operators. 
The Moyal phase will now depend
on the parameter of the metric which will play the role of the
non-commuting parameter. This is the main result of this Paper,
it is possible to have a non-commutative string theory also
when one has a metric of a particular form and no $B$-field.
The metric background considered admits a NCOS limit
in which the closed strings decouple.

When the Dirichlet spatial direction is compactified 
the theory becomes T-dual to the NCOS considered in~\cite{Klebanov:2000pp}
and there are finite energy closed string modes with a
positive discrete momentum.
In this case we give an explanation for the origin of noncommutativity
that differs form the one given 
in \cite{Danielsson:2000gi}, where is suggested that the Moyal
phase in the T-dual picture emerges as a consequence of a large 
boost of the system. In our approach the appearance of the 
Moyal phase is a consequence of the Buscher rules  
and of the correct definition of the propagators and of 
the vertex operators in the dual theory.

In section 2 we will perform a canonical analysis for an
open bosonic string in the presence of a
metric in the $0$ and $1$ directions, when the direction $1$ 
has Dirichlet boundary conditions. 
Namely we consider an open string
propagating in the background of a D24-brane. 
Here we will derive the Virasoro generators and the energy
spectrum.

In section 3 we will compute
the propagators in this theory and define the vertex operators.
Taking the operator product expansion of two vertex operators
we shall then compute the Moyal phase. 

Section 4 is devoted to the analysis of the situation in which both a metric
and a $B$-field are present. We shall show that the Moyal phase
depends either on the metric moduli or on $B$ depending on the boundary
conditions.

In section 5 we shall consider the situation in which the direction $1$
is compactified on a circle of radius $R$ and discuss T-duality.

We shall then revisit the arguments of \cite{Klebanov:2000pp}, 
consider the T-duality relations between different 
backgrounds and show in which way the NCOS limit for each 
of them has to be taken. The relation with the DLCQ limit 
of closed string is discussed. We then consider the case 
when both the metric and the background $B$ field are present. 

In section 7, we consider the high temperature behavior
of the various background we examined, showing 
that the Hagedorn temperature for open strings 
depends on the background moduli space. This dependence
is the same as the one found for closed string
when there is a compactified direction~\cite{Grignani:2001ik}.

\section{Canonical Analysis} 

We will begin by examining the effect of a certain simple
background on the canonical analysis of theories 
of open bosonic strings. This background is a spacetime metric of the form 
\be g_{\mu\nu}= 
\left( \matrix{ -1+A^2 & -A & 0 & \ldots \cr -A & 1 & 0 & \ldots 
\cr 0 & 0 & 1 & \ldots \cr \ldots & \ldots & \ldots & \ldots \cr } 
\right) 
\label{metric} 
\ee 
where $A$ is a constant. To keep the $g_{00}$ component of the 
metric time-like $A$ must be less than $1$, 1 is a 
critical value for $A$.  
The action is 
\be 
S=-\frac{1}{4\pi\alpha'}\int d^2\sigma 
g_{\mu\nu} \partial^\alpha X^\mu\partial_\alpha X^\nu 
\label{action} 
\ee
The boundary conditions we assume are
\bea 
\d_\tau X^1 \mid_{\sigma=0,\pi} & = & 0 \nn \\ 
\left[(1-A^2) \d_\sigma X^0  + A \d_\sigma X^1 
\right]\mid_{\sigma=0,\pi} & = & 0 \nn \\ 
\d_\sigma X^a \mid_{\sigma=0,\pi} & = & 
0 ~~~~~~~~~~~
a=2,...,26
\label{bc} 
\eea 
namely the direction $1$ is Dirichlet and the
transverse directions are Neumann.
This background is obtained by a duality transformation 
\cite{Buscher:1988qj,Buscher:1987sk,Giveon:1994fu} 
from a theory in which the space-time metric is Minkowskian,
there is a non-zero Neveu-Schwarz $B$ field 
\be
S=-\frac{1}{4\pi\alpha'}\int d^2\sigma \left[ \eta_{\mu\nu} 
\partial^\alpha X^\mu\partial_\alpha X^\nu - \varepsilon^{\alpha 
\beta}B_{\mu\nu} \partial_\alpha X^\mu\partial_\beta X^\nu \right] 
\label{actionB} \ee 
with
\be B_{\mu\nu}= \left( \matrix{ 0 & B & 0 
& \ldots \cr -B & 0 & 0 & \ldots \cr 0 & 0 & 0 & \ldots \cr \ldots 
& \ldots & \ldots & \ldots \cr } \right)  
\label{bfield}
\ee
and the boundary conditions are
\bea 
\left[\d_\sigma X^1 +B\d_\tau X^0\right]\mid_{\sigma=0,\pi} & = & 0 \nn \\ 
\left[\d_\sigma X^0  + B \d_\tau X^1 
\right]\mid_{\sigma=0,\pi} & = & 0 \nn \\ 
\d_\sigma X^a \mid_{\sigma=0,\pi} & = & 
0 ~~~~~~~~~~~
a=2,...,26
\label{bc1}
\eea 
The Buscher rules for this system in fact are
\be
g'_{00}=-1+B^2~,~~~~ g'_{01}=B~,~~~~g'_{11}=1~,~~~~B'_{\mu\nu}=0
\ee
and lead to a background of the type (\ref{metric}) with $A\to -B$.

The variation of the action (\ref{action}) yields the equations of motion 
and constraints
\be
\partial_{\alpha}\partial^{\alpha}X^{\mu}=0
~~~,~~~g_{\mu \nu}\partial_{\tau}X^{\mu}
\partial_{\sigma}X^{\nu}=0
\label{eom}
\ee
The solutions of (\ref{eom})
in the background (\ref{metric}) with the boundary conditions
(\ref{bc}) for the $0$ and $1$ directions are 
\bea 
X^0 (\tau,\sigma) & = & x^0+\sqrt{2\alpha'}
\left( \alpha_0^0+\frac {A} {1-A^2}\alpha_0^1 \right)\tau + 
\sqrt{2\alpha'} \frac {A} {1-A^2} \alpha_0^1 \sigma \nn \\ 
&+& i \sqrt{2\alpha'} \sum_{n\neq 0} \left( \frac {\alpha_n^0} {n} 
e^{-in\tau}\cos n\sigma + \frac {A} {1-A^2} \frac {\alpha_n^1} {n} 
e^{-in(\tau +\sigma)} \right) \nn \\ 
X^1 (\tau,\sigma) & = & x^1 - \sqrt{2\alpha'} \alpha_0^1 \sigma -
\sqrt{2\alpha'} \sum_{n\neq 0} 
\left( \frac {\alpha_n^1} {n} e^{-in\tau}\sin n\sigma \right) 
\label{expansion} 
\eea 
whereas the transverse directions have the usual expansions for Neumann
coordinates. The conjugate momenta read 
\bea 
\Pi_0 & = & \frac {1} {\pi} \frac {1} 
{\sqrt{2\alpha'}} \left[-(1-A^2) \left( \alpha_0^0 + 
\frac {A} {1-A^2} \alpha_0^1 \right) 
+  \sum_{n\neq 0} e^{-in\tau} \cos n\sigma 
\left(-(1-A^2) \alpha_n^0-A\alpha_n^1 \right) \right] \nn \\ 
\Pi_1 & = & \frac {1} {\pi} \frac {1} 
{\sqrt{2\alpha'}} \left[-A\alpha_0^0 - \frac {A^2} {1-A^2} 
\alpha_0^1 + \sum_{n\neq 0} e^{-in\tau} 
\cos n\sigma \left(-A\alpha_n^o - \frac {A^2} {1-A^2}\alpha_n^1 
\right) \right. \nn \\ 
& & \left.  + \frac {i} {1-A^2} \sum_{n\neq 0} e^{-in\tau} 
\sin n\sigma \alpha_n^1 \right] 
\label{dens.momenta} 
\eea 
The total momentum in the direction $1$, which is
obtained by integrating in $\sigma$ the expression for $\Pi_1$,
is not conserved. This is due to the fact 
that the presence of the $D$-brane breaks 
translational invariance along the direction 1, so it is not 
possible to define consistently a momentum which is canonically 
conjugate to the center of mass position $x^1$. 
The total momentum $P_0$ is instead defined in the usual way 
\be
P_0 =  
\int_0^\pi d\sigma \Pi_0(\tau,\sigma) = -\frac {1} 
{\sqrt{2\alpha'}} \left( (1-A^2) \alpha_0^0 +A\alpha_0^1 \right) 
\label{can.momenta} 
\ee
The equal time commutation relations
have the standard free field form 
\bea
\left[ X^\mu(\t,\s),\Pi_\nu(\t,\s') \right] & = & 
i\delta^\mu_{~\nu}\delta(\s-\s') \nn \\
\left[ X^\mu(\t,\s),X^\nu(\t,\s')\right] & = & 
\left[\Pi^\mu(\t,\s),\Pi^\nu(\t,\s')\right] = 0
\label{ccr}
\eea
These are consistent with the boundary conditions (\ref{bc})
and there seems to be no space/time noncommutativity.
What happens here is quite different to the dual case
\cite{Chu:1998qz}, where the boundary conditions (\ref{bc1})
are inconsistent with the
canonical commutation relations and noncommutativity appears 
already at this stage. 

In order to obtain the relations (\ref{ccr}), 
the oscillation modes must satisfy the
standard commutation relations 
\be
\left[\a_n^\mu, \a_m^\nu\right] = n\delta_{n+m,0}g^{\mu\nu}~,~~~~~n,m\ne 0
\label{modescomm1}
\ee
and the commutators between the
total momenta $P_0$ and $P_a$ $a=2,\dots,26$ 
and the center of mass modes $x^\mu$ are as expected 
\be
\left[x^a,P_b\right]  =  i\delta^a_{~b}~,~~~~~~ 
\left[x^0,P_0\right]  =  i~,~~~~~~
\left[x^1,P_0\right]  = 0 
\label{modescomm}
\ee 

The commutation relations (\ref{modescomm1}), (\ref{modescomm})
are just enough to satisfy the canonical commutators,  
but something seems to be missing. In fact in these commutation relations
there is no information about the degree of freedom related to the zero mode
of the direction 1. This operator
should be proportional to $\a_0^1$. 
To construct the complete Fock space of the string,
instead of the canonical momentum $P_1$, it is then necessary to introduce 
the operator
\be
L^1=\frac{1}{\sqrt{2\a'}}\a_0^1
\label{diffp1}
\ee 
and the generic state of the string should be represented as 
$|N,k_0,l^1,{\vec k}>$
where $l^1$ is the eigenvalue of $L^1$.
It is easy to see that this operator commutes with the other canonical
momenta. The physical interpretation of this operator becomes clear
when the direction $1$ is compactified. Its eigenvalue is
proportional  to the winding
number of the open string around the compact direction.
We shall come back to this in section 5.

To construct the vertex operators it is necessary to find
the variable $Q^1$ which is canonically 
conjugate to $L_1$, $i.e.$ 
$\left[Q^1,L_1\right]=i$. For this purpose
we invoke the doubling formalism to solve 
the equations of motion 
(\ref{eom}). In fact the most general solution of (\ref{eom}) is the sum 
of the left, $X^1_L$, and right, $X^1_R$, mover modes 
\bea 
X^1_L (\tau - \sigma) & = & x^1_L+\sqrt{\frac{\alpha'}{2}}
\a_0^1 (\t -  \s)   + \mbox{ osc.} \nn \\ 
X^1_R (\tau + \sigma) & = & x^1_R + \sqrt{\frac{\alpha'}{2}} 
\tilde{\alpha}_0^1 (\t + \s) + \mbox{ osc.}
\label{expansion2} 
\eea 
thus leading to the definition of a left and a right momentum
$p^1_L=\sqrt{\frac{2}{\a'}}\a_0^1$ and 
$p^1_R=\sqrt{\frac{2}{\a'}}\tilde{\alpha}_0^1$.
Taking into account the boundary conditions (\ref{bc}) it is easy to see that
$\a_0^1 =-\tilde{\alpha}_0^1 $, so that the operator (\ref{diffp1}) 
is given by the difference $L^1 = \left(p^1_L - p^1_R\right)/2$.
From this we can argue \cite{Horava:1989ga,Sakamoto:1989ig} 
that the variable canonically conjugate to $L^1$ is given by
the difference of the constant modes of the left and right 
expansion (\ref{expansion2})
\be
Q^1 = x_L^1-x_R^1 
\label{newcommrel}
\ee

In what follows we shall also need the energy spectrum. This can be easily
derived from the Virasoro generators
\be
L_m=\sum_{n\in Z}g_{\mu\nu}\alpha^{\mu}_{m-n}\alpha^{\nu}_{n}
\ee
and the condition $L_0-1=0$. This yields 
\be
k_0=
\sqrt{\left(l^1\right)^2 +\left(1-A^2\right)
\left[{\vec{k}}^2 +\frac{\left(N-1\right)}{\alpha'}\right]}
\label{2spectrum}
\ee 
where $k_0$ and $l^1$ are the eigenvalues of $P_0$ and 
$L^1$, respectively. $\vec k$ is the
transverse momentum and $N=\sum_{n=1}^\infty 
\alpha_{-n}\cdot\alpha_n$ with a standard 
notation for oscillators~\cite{Green:1987sp}.

\section{Evaluation of the Moyal Phase}

In this section we compute the product of two tachyon vertex
operators and show how the space/time noncommutativity manifests itself:
the product of two vertex operators will give rise to a
Moyal phase, namely it can be interpreted as a Moyal $\ast$ product.
We will work at tree level in string theory and represent
the string worldsheet as the upper half plane using the convention of
ref.\cite{Polchinski:rq}.

Consider the
propagator of the theory in the background (\ref{metric}) and with
boundary conditions (\ref{bc})
\be 
<X^\mu(z_1) X^\nu (z_2)> =  -\frac{\a'}{2}\left[g^{\mu\nu} \log|z_1-z_2|^2 -
A^{\mu\nu} \log|z_1-\bar{z_2}|^2  +D^{\mu \nu} \right]
\label{propagators}
\ee
where
\bea
g^{\mu\nu}  =  \left( \matrix{-1 & -A & 0 & \ldots 
\cr -A & 1-A^2 & 0 & \ldots \cr 0 & 0 & 1 & \ldots \cr \ldots &  
\ldots & \ldots & \ldots \cr }\right)~,~~~~~
A^{\mu\nu}  =  \left( \matrix{\frac{1+A^2}{1-A^2} & -A & 0 & \ldots 
\cr -A & 1-A^2 & 0 & \ldots \cr 0 & 0 & 0 & \ldots \cr \ldots &  
\ldots & \ldots & \ldots \cr }\right)
\label{iga}
\eea
The constants $D^{\mu \nu}$ are independent on $z_1$ and $z_2$ and can
be set to a convenient value.
The propagator (\ref{propagators}) is symmetric
so that the boundary propagator will not contain a 
noncommutativity parameter.
By interpreting time ordering as operator ordering the commutator
$[X^0(\tau),X^1(\tau)]$ will just vanish.
The antisymmetric contribution responsible for the
Moyal phase in the NCOS theories~\cite{Seiberg:1999vs} seems to be absent.

We must now find an appropriate expression for the vertex operators. 
They should posses the same symmetry of
the correspondent states both under translation and under the symmetry
generated by the operator (\ref{diffp1}), which can be viewed as  a 
translation of the
left and right parts of $X^1$ in opposite directions. This is
consistent with the fact that the dual symmetry reverses the sign of the right 
part
of the coordinate: $\tilde{X^1}(z,\bar{z})=X^1_L(z)-X^1_R(\bar{z})$.  
Since the operator $L_1$ commutes with the other operators that 
generate the Fock space, its eigenvalue is preserved throughout any
interaction and the vertex operator should depend on it.
Consequently, for the 
tachyon vertex operator one can write~\cite{Horava:1989ga}
\be
\mathcal{V}_{k_0,w,\vec k}=
:e^{ik_0X^0(z,\bar{z})+il_1(X^1_L(z)-X^1_R(\bar{z}))+ik_a X^a(z,\bar{z})}:
\label{vo}
\ee
where $k_0$, $l_1$ and $k_a$ are the eigenvalue of
$P_0$, $L_1=\a_0^1/2\a'(1-A^2)$ and of the transverse momenta $P_a$.

The world-sheet integral of the vertex operator 
must be conformally invariant. This implies
that (\ref{vo}) must be a tensor of weight ($1/2$,$1/2$). 
By a simple OPE computation,
it is possible to show that the vertex (\ref{vo}) is a tensor of weight
$$
h=\bar h =
- \frac{\alpha'}{2\left(1-A^2\right)}\left[k_0^2- (l^1)^2
- \left(1-A^2\right) {\vec k}^2\right]
$$
This gives for the tachyon
\be
k_0=
\sqrt{(l^1)^2 + (1-A^2)
\left({\vec{k}}^2 -\frac{1}{\alpha'}\right)}
\label{vos}
\ee 
which agrees with the expression derived from the
constraint $L_0 -1=0$. This again confirms the validity of the
expression (\ref{vo}) for the vertex operator.

Since the vertex (\ref{vo})
depends on the holomorphic and anti-holomorphic parts
of $X^1(z_1,\bar{z}_1)=X^1_L(z)+X^1_R(\bar{z})$ 
it is useful to explicitly write the propagators of 
these modes separately. These can be obtained from
(\ref{propagators}) by keeping into account that 
$X_L(z)$ is holomorphic and $X_R(\bar{z})$ anti-holomorphic.
The relevant propagators are
\bea 
&&<X^0(z_1) X^1_L(z_2)> =  \frac{\a'}{2}A\log \frac{z_1-z_2}
{\bar{z_1}-z_2}\cr
&&<X^0(z_1) X^1_R(\bar{z}_2)> =  
\frac{\a'}{2}A\log \frac{\bar{z}_1-\bar{z}_2}
{z_1-\bar{z}_2}\cr
&&<X^1_L(z_1) X^1_L (z_2)> = -\frac{\a'}{2}\left(1-A^2\right)
\log\left( z_1-z_2 \right)\cr
&&<X^1_R(\bar{z}_1) X^1_R(\bar{z}_2)> = -\frac{\a'}{2}\left(1-A^2\right)
\log\left( \bar{z}_1-\bar{z}_2 \right)\cr
&&<X^1_L(z_1) X^1_R (\bar{z}_2)> = \frac{\a'}{2}\left(1-A^2\right)
\log \left(z_1-\bar{z}_2 \right)
\label{prop}
\eea
which again are defined up to a constant.

Consider now the OPE for the product of two vertex
operators inserted on the boundary of the worldsheet, 
$\mathcal{V}_1(\tau)\mathcal{V}_2(0)$.
To evaluate this OPE consider the propagators (\ref{propagators})
and (\ref{prop})
at the boundary points $\tau$ and $0$. We have
\bea
&&<X^0(\tau) X^0(0)>=\frac{\alpha'}{1-A^2}\log\tau^2 \cr
&&<X^0(\tau) \left(X^1_L(0)-X^1_R(0)\right)> =-i\alpha' \pi A 
\epsilon (\tau) \cr
&&<\left(X^1_L(\tau)-X^1_R(\tau)\right)\left(X^1_L(0)-X^1_R(0)\right)>=
\alpha' \left(1-A^2\right)\log\tau^2
\label{bp}
\eea
where $\epsilon(\tau)$ is the function that is $1$ or $-1$ for positive
or negative $\tau$ and the constant $D^{\mu \nu}$ has been set to
a convenient value.
With the boundary propagators (\ref{bp}) the product of two normal ordered
tachyon vertex operators satisfy (we ignore the transverse coordinates)
\bea
&&:e^{ik_0X^0(\tau) + i l_1 \left(X^1_L(\tau)-
X^1_R(\tau)\right)}:
:e^{ik'_0X^0(0) +
il'_1 \left(X^1_L(0)-X^1_R(0)\right)}: \cr
&&= \exp\left\{-\frac{\alpha'}{1-A^2} \left(k_0k'_0 
- l^1l'^1 \right)\log \tau^2
+i\pi \alpha'\epsilon(\tau) \frac{A}{1-A^2}\left(k_0l'^1 -k'_0l^1\right)\right\}\cr
&&
:e^{ik_0X^0(\tau)+il_1(X^1_L(\tau)-X^1_R(\tau))
+ik'_0X^0(0)+il_1(X^1_L(0)-X^1_R(0))}:
\label{OPE1}
\eea
where $l^1=(1-A^2)l_1$ is the quantity dual to the momentum in the direction 1,
$k_1$, of the dual theory.
The form of the OPE (\ref{OPE1}) is precisely dual to
the well known expression derived in~\cite{Seiberg:1999vs}
and in the $\a'\to 0$ limit defines a Moyal product
with non commutative parameter given by 
\be
\theta=2\pi \a'\frac{A}{(1-A^2)}
\ee
The Moyal phase is
\be
\exp{\left[i\pi \a'\frac{A}{(1-A^2)}
\left(k_0l'^1 -k'_0l^1 \right)\right]} 
\label{phasenew} 
\ee
In (\ref{OPE1}) the short distance singularity is governed
by the ``open string metric''
\bea
G^{\mu\nu}  =  \frac{1}{1-A^2}\left( \matrix{-1 & 0 & 0 & \ldots 
\cr 0 & 1 & 0 & \ldots \cr 0 & 0 & 1-A^2& \ldots \cr \ldots &  
\ldots & \ldots & \ldots \cr }\right)
\eea
and determines the anomalous dimension
of the vertex operators.

Now let $A=1-\epsilon /2$. We can define a NCOS type limit as 
\be
\epsilon \to 0,~~~{\rm with}~~ \alpha'=\alpha'_e \epsilon ,~~
{\vec K} =\vec k \epsilon ~;~~~\alpha'_e ~~{\rm fixed}
\ee
In this limit the open string spectrum (\ref{2spectrum}) becomes
\be
k_0=
\sqrt{\left(l^1\right)^2 +
\left[ {\vec{K}}^2 +\frac{\left(N-1\right)}{\alpha'_e}\right]}
\ee 
and remains finite. As in~\cite{Seiberg:2000ms}
the closed string spectrum instead diverges and closed strings
decouple.
The anomalous dimension of the vertex operator could be ignored
and the Moyal phase becomes
\be
\exp{\left[i\pi \a'_e\left(k_0l'^1 -k'_0l^1 \right)\right]} 
\ee

\section{Inclusion of the $B$ Field}

Next we may add a constant Neveu-Schwarz $B$-field to the action.
It becomes
\be
S=-\frac{1}{4\pi\alpha'}\int d^2\sigma \left[ g_{\mu\nu} 
\partial^\alpha X^\mu\partial_\alpha X^\nu -
\varepsilon^{\alpha \beta}B_{\mu\nu} \partial_\alpha 
X^\mu\partial_\beta X^\nu \right] 
\label{actionB1} 
\ee 
The $B$-dependent term in the action (\ref{actionB1}) is a total 
derivative, so it does not affect the equations of motion. 
Since the direction $1$ is Dirichlet the boundary conditions 
are still given by (\ref{bc}).
Consequently, also the propagators are left unchanged. However,
the presence of the $B$ field affects the energy spectrum 
by shifting its value by a constant:     
\be
k_0=Bl^1 + 
\sqrt{(l^1)^2 +\left(1-A^2\right)\left[\vec{k}^2 +
\frac{\left(N-1\right)}{\alpha'}\right]}
\label{spectrumb}
\ee 

It is interesting to notice that contrary to
what happens in NCOS theories, in this case the antisymmetric
tensor field $B_{\mu \nu}$ does not participate to the Moyal phase 
(\ref{phasenew}). $B$ affects only $P_0$ as in (\ref{spectrumb}),
and, being the propagators unchanged, the phase keeps the form 
(\ref{phasenew}). In fact, since the eigenvalue $k_0$ is shifted
by $Bl^1$, the $B$ dependent terms cancel.

In the same way it is possible to show that starting with the dual
theory with both the metric and the $B$ field,
even if the metric modifies the energy spectrum of the theory, 
it does not affect the Moyal phase.
Let us consider this case in more detail.
The action is again (\ref{actionB1}) but the boundary conditions now are 
\bea 
\left[g_{00}\d_\sigma X^0 +g_{01}\d_\sigma X^1 -B\d_\tau X^1\right]
\mid_{\sigma=0,\pi} & = & 0 \nn \\ 
\left[g_{10}\d_\sigma X^0 +g_{11}\d_\sigma X^1\d_\sigma X^0  + B \d_\tau X^0 
\right]\mid_{\sigma=0,\pi} & = & 0 \nn \\ 
\d_\sigma X^a \mid_{\sigma=0,\pi} & = & 
0 ~~~~~~~~~~~
a=2,...,26
\label{bcab}
\eea
The equations of motion 
and constraints are given by (\ref{eom}) and,
taking into account the boundary conditions (\ref{bcab}), have
the solutions
\bea
X^0 (\tau,\sigma) & = & x^0+ 2\alpha' p^0 \tau +
2\alpha'ABp^0 \sigma -2\alpha'Bp^1 \sigma+  \cr 
&+& \sqrt{2\alpha'} \sum_{n\neq 0} \frac{e^{-in\tau}}{n}
\left( i\alpha_n^0 \cos n\sigma + AB\alpha_n^0 \sin n\sigma 
-B\alpha_n^1 \sin n\sigma \right) \cr 
X^1 (\tau,\sigma) & = & x^1 + 2\alpha' p^1 \tau -
2\alpha'ABp^1 \sigma +2\alpha'\left(A^2 -1\right)Bp^0 \sigma  \cr 
&+& \sqrt{2\alpha'} \sum_{n\neq 0} \frac{e^{-in\tau}}{n}
\left[ i\alpha_n^1 \cos n\sigma -AB\alpha_n^1 \sin n\sigma 
+\left(A^2 -1\right)B\alpha_n^0 \sin n\sigma \right]~~~
\label{expansionab} 
\eea
and the usual expansions hold for the transverse Neumann coordinates.
Following~\cite{Chu:1998qz}
the commutation relations of the modes in (\ref{expansionab}) are
\be
\left[x^\mu,x^\nu\right]=i \theta^{\mu\nu}~,~~~
\left[x^\mu,p^\nu\right]=i {M^{-1}}^{\mu\nu}~,~~~
\left[\alpha_m^\mu,\alpha_n^\nu\right]=im\delta_{m+n,0} {M^{-1}}^{\mu\nu}
\ee
where 
\be
M_{\mu\nu}= g_{\mu \nu}-B_{\mu \rho}B^{\rho}_{\ \nu}~,~~~~~
\theta^{\mu\nu}=2\pi\a'\frac{B^{\mu\nu}}{1-B^2}
\label{mij} 
\ee

From the action (\ref{actionB1}) the canonical momentum is
\be
P^{\mu}=\frac{1}{2\pi \alpha'}\int_{0}^{\pi}d\sigma
\left(\partial_{\tau}X^{\mu}
+\partial_{\sigma}X^{\nu}B_{\nu}^{\ \mu}\right)
=p^{\nu}M_{\nu}^{\ \mu}
\label{pmu}
\ee

The momentum $P^{\mu}$ is conjugate in the usual sense
to the center
of mass coordinate $x^{\mu}_{\rm c. m.}$ defined as
$$
x^{\mu}_{\rm c. m.}=\frac{1}{\pi }\int_{0}^{\pi}d\sigma
X^{\mu}\left(\tau, \sigma \right)
$$
By imposing the equation (\ref{eom}) on the physical
states one can derive the Virasoro generators \cite{Chu:1998qz}
\be
L_m=\sum_{n\in Z}M_{\mu\nu}\alpha^{\mu}_{m-n}\alpha^{\nu}_{n}
\ee
so that the energy spectrum reads
\be
k_0= -Ak_1 +
\sqrt{k_1^2 +\left(1-B^2\right)
\left[{\vec{k}}^2 +\frac{\left(N-1\right)}{\alpha'}\right]}
\label{aspectrum}
\ee
This energy spectrum is dual to (\ref{spectrumb}),
it becomes (\ref{spectrumb}) when $A\leftrightarrow -B$ 
and the momentum along the  direction 1 is substituted
with the eigenvalue of the operator $L^1$.

The vertex operator in this case has the usual form
\be
\mathcal{V}_{k_\mu}=
:e^{ik_\mu X^\mu(z,\bar{z})}:
\label{vos1}
\ee
With the  boundary conditions (\ref{bcab}) the
propagators read
\bea
<X^\mu(z_1) X^\nu (z_2)> &=& -\frac{\a'}{2}\left[g^{\mu\nu} 
\log|z_1-z_2|^2-g^{\mu\nu}\log|z_1-\bar{z}_2|^2\right. \cr
&&\left.+2G^{\mu \nu}\log|z_1-\bar{z}_2|^2+
\frac{1}{\pi \alpha'}\theta^{\mu \nu}\log|z_1-\bar{z}_2|^2+D^{\mu \nu}\right]
\label{propagators1}
\eea
where $g^{\mu\nu}$ is the metric (\ref{iga}) and 
$$
G_{\mu \nu}=g_{\mu \nu}-\left(B g^{-1} B\right)_{\mu\nu}
$$
is the open string metric.

In the OPE of two normal ordered vertex operators a Moyal phase with 
the form 
\be
\exp{\left[-{i\pi\alpha'}\frac{B}{\left(1-B^2\right)}
\left(k_0 k_1' -k'_0 k_1\right)\right]} 
\label{phasenew1} 
\ee
appears. 
It does not depend on $A$ and is dual to (\ref{phasenew}).

\section{The Compact Case}

In this section we will compactify
the direction $1$.
\be
x^1 \sim x^1 + 2\pi R 
\label{compact} 
\ee
The duality symmetry discussed in the previous sections becomes now
T-duality. The origin of non-commutativity in T-dual NCOS
has raised some discussions~\cite{Danielsson:2000gi,Maharana:2000fc},
here we give a different interpretation.

Consider first the case described by the action (\ref{action})
with boundary conditions (\ref{bc}). 
Being the direction 1 compactified 
we must require 
\be\alpha_0^1 = \sqrt{\frac{2}{\alpha'}}Rw
\label{winding}
\ee 
where $w$ is an integer. 
As a matter of fact an open string with Dirichlet boundary conditions along
a compact dimension can have non-trivial winding modes.
Since the ends of the string are tied to the D-brane it cannot unwrap.
On the other hand the Dirichlet string does not have Kaluza Klein
momenta. 

We may now clarify the physical meaning of the operator $L^1$. 
When the direction $1$ is compactified, 
it becomes exactly the operator associated with the
winding modes of the string. So, in the decompactified limit we find a sort of
``continuum winding'', described by the eigenvalue $l^1$.
All the computations above are unchanged, but one must set
\be
l^1=\frac{wR}{\a'}
\ee
In particular the Moyal phase becomes
\be
\exp{\left[i\pi R\frac{A}{\left(1-A^2\right)}
\left(k_0w' -k'_0w \right)\right]} 
\label{phasenew2} 
\ee
This expression is manifestly T-dual to (\ref{phasenew1}), where,
due to the compactification, $k_1$ is quantized as $k_1=m/R$.

\section{Closed Strings} 

In \cite{Klebanov:2000pp} it was argued that when there is a 
compactified direction, wound states of closed strings do not 
decouple from the spectrum in the NCOS limit. These closed string 
states were used to construct Wound String theory in 
\cite{Danielsson:2000gi} and Non-Relativistic Closed Strings 
\cite{Gomis:2000bd}. In this section we will see how these 
arguments apply to our situation. For closed strings the canonical 
momenta in the compactified direction is quantized 
\be 
k_1=\frac{m}{R},~~~~~~~~~~m\in Z \label{p1quantizz} 
\ee 
the energy 
spectrum for closed string when only the metric (\ref{metric}) is 
present reads 
\be k_0= -\frac{Am}{R} + \sqrt{ 
\left(\frac{wR}{\alpha'}\right)^2 
+\left(\frac{m}{R}\right)^2+ {\vec k}^2+ 
\frac{2}{\alpha'}(N+\tilde N-2) } \label{SpclosedA} 
\ee 
where 
$\vec k$ denotes the transverse momentum, $N=\sum_{n=1}^\infty 
\alpha_{-n}\cdot\alpha_n$ and $\widetilde{N}=\sum_{n=1}^\infty 
\widetilde{\alpha}_{-n} \cdot \widetilde{\alpha}_n$. 
Let 
$A\thicksim1-\varepsilon/2$ and define the relevant constant for 
the limiting process (notice that in the usual description $R$ is 
held fixed) 
\be \alpha'_e = \frac{\alpha'}{\varepsilon}, ~~~~~~ 
{\vec k}^2 = \frac{{\vec K}^2}{\varepsilon},~~~~~~ R_e = 
\frac{R}{\varepsilon} \label{NCOSconst} 
\ee 
The NCOS limit 
\cite{Seiberg:2000ms,Gopakumar:2000na} consists of taking 
$\varepsilon\rightarrow0$, keeping the constants (\ref{NCOSconst}) 
fixed. The spectrum of closed string diverges in this limit, 
unless $m>0$, and in this case we obtain: 
\be k_0= 
\frac{m}{2R_e}+\frac{R_e}{2m}\vec K^2+ \frac{R_e}{\alpha'_e m} 
(N+\tilde N-2) 
\label{NCOSspecA} 
\ee 
The result that $k_1$ must be 
positive is an indication of the T-duality relation between NCOS 
theories and the DLCQ strings \cite{Klebanov:2000pp, Danielsson:2000gi}. 
In fact, defining $k^+=\sqrt{2}k^0 - k^-$ and 
$k^-=m/(\sqrt{2} R_e)$ we obtain exactly the DLCQ closed string 
spectrum \cite{Grignani:1999sp,Grignani:2000zm}
 \be 
k^+=\frac{N+\tilde N-2}{\alpha'_e k^-}+\frac{\vec K^2}{2k^-} 
\label{bdlcq} 
\ee 
Notice 
also that the expression (\ref{NCOSspecA}) is T-dual to the 
spectrum obtained in \cite{Danielsson:2000gi} with the parameter 
(\ref{NCOSconst}) of the NCOS theory and taking into account that in 
the dual situation $R=R_e$. 

By adding the $B$ field one changes the spectrum according to 
\be k_0= 
\frac{BRw}{\alpha'}-\frac{Am}{R} +\sqrt{ 
\left(\frac{wR}{\alpha'}\right)^2 
+\left(\frac{m}{R}\right)^2+\vec k^2+ 
\frac{2}{\alpha'}(N+\tilde N-2) } 
\label{SpclosedB} 
\ee 
It is now possible to take the NCOS 
limit by letting $B=1-\varepsilon/2$ and by keeping $\alpha'_e$, the 
rescaled transverse momentum $\vec K^2$ and the radius constant as 
in \cite{Danielsson:2000gi}, thus obtaining
\be 
k_0= -\frac{Am}{R} 
+ \frac{wR}{2\alpha'_e}+\frac{\alpha'_e}{2wR}\vec K^2+ 
\frac{N+\tilde N-2}{wR} 
\label{NCOSspecB1} 
\ee 
where $w>0$. Otherwise one can put
$A=1-\varepsilon/2$ and by keeping constant the parameter 
(\ref{NCOSconst}) in the limit $\epsilon\to 0$ the spectrum reads
\be 
k_0= \frac{BR_ew}{\alpha'_e} + \frac{m}{2R_e} + 
\frac{R_e}{2m}\vec K^2+ \frac{R_e}{\alpha'_e m} (N+\tilde N-2) 
\label{NCOSspecB} 
\ee 
with $m>0$. Of course this two cases are related 
by T-duality, with $R=R_e$ in the first case, and from both it is 
easy to derive the DLCQ spectrum for closed string in the presence of 
a background $B$ or $A$ field.

\section{The Hagedorn Transition}

To better understand the physics of the strings in the 
background we are considering it is useful to subject them
to extreme conditions by heating them up to high temperatures.
For conventional superstring theory this has been extensively studied
(see for instance 
\cite{Bowick:az,Alvarez:1986sj,Atick:1988si,Deo:1989bv}).
In the presence of spacetime backgrounds the Hagedorn temperature
was studied for example in refs.
\cite{Ferrer:1990na,Tseytlin:1998kw,Grignani:2001hb,Grignani:2001ik}.
The systems studied in these papers are gravitating so that it
is difficult to study their thermodynamics. Nevertheless the Hagedorn
transition has been interpreted as a first order phase transition
\cite{Atick:1988si}.
Since the NCOS are decoupled from gravity their 
thermodynamics can be analyzed in detail and
the phase diagram has an extremely rich 
structure~\cite{Gubser:2000mf,Barbon:2001tm,Chan:2001gs}. It also has
gauge theory analogs as shown in
\cite{Kogan:2001px}.
The NCOS transition becomes of second order and can be studied
in the context of weakly coupled string theory.
However, when a direction is compactified wrapped states of
closed strings do not decouple from the spectrum
in the NCOS limit~\cite{Klebanov:2000pp}.
One would expect that, in the limit as the 
compactification radius is large, the wrapped closed strings would couple 
more and more weakly and in the
infinite, de-compactified limit they would disappear from the spectrum.  
Indeed their energies do go to infinity.  However it was shown in
\cite{Grignani:2001ik} that their Hagedorn temperature remains, 
that is, no matter how 
large that radius is, they still participate in the Hagedorn transition.  
It was argued in~\cite{Grignani:2001ik} that the closed string 
Hagedorn behavior makes it a first order transition again.

In this section we will investigate the thermodynamic 
properties of the open string sector in the backgrounds considered
above showing that the Hagedorn temperature is modified by the
presence of the backgrounds.
Finally, we will compactify the direction $1$ and 
we will show that for the open string sector the dependence
of the Hagedorn temperature on the
background moduli has the same non-extensive (radius independent) 
behavior as that of the
closed string sector~\cite{Grignani:2001ik}.
The two sectors undergo a phase transition at the same time.

Consider first the case in which the background is only metric
(\ref{metric}) with boundary conditions (\ref{bc}).

The free energy of a gas of relativistic Bose particles is
\be
F=\frac{1}{\beta}{\rm Tr}\ln\left( 1-e^{-\beta
P_0}\right)= -\sum_{n=1}^\infty \frac{1}{n\beta}{\rm Tr} e^{-n
\beta P_0}
\label{particle}
\ee
As well known,
(\ref{particle}) can be used to derive the bosonic string free 
energy at one loop, by the standard procedure of computing the sum
of free energies of the particles in the string spectrum.
The energy spectrum is given by equation (\ref{2spectrum}).

To obtain the free energy of the bosonic string we use the
integral identity
$$ 
\int_0^\infty dt e^{-x t^2-y/{t^2}}=
\frac{1}{2}\sqrt{\frac{\pi}{x}}e^{-2\sqrt{xy}} 
$$ where
$$
t^2=1/\tau_2,~~~ x=\frac{n^2\beta^2}{4\pi \alpha'}\left(1-A^2\right), ~~~
y=\pi \alpha' \left[\frac{(l^1)^2}
{\left(1-A^2\right)}
+{\vec P}^{~2}+ \frac{1}{\alpha'}(N-1)\right]
$$
Then the free energy reads
\be
F=-\sum_{n=1}^{\infty} \int_0^{\infty} \frac{d\tau_2}{\tau_2}
\frac{1}{(4\pi^2\alpha'\tau_2)^{\frac{D}{2}}}
\left|\eta\left(\frac{\tau_2}{2}\right)\right|^{-24}
e^{-\frac{\beta^2 n^2}{4\pi\alpha'\tau_2}\left(1-A^2\right)
-\pi\alpha'\tau_2\frac{(l^1)^2}
{\left(1-A^2\right)}}
\label{free}
\ee
The temperature independent $n=0$ term
gives the vacuum energy, $i.e.$ the cosmological constant
contribution, the other terms give the relevant thermodynamic
potential.
The Hagedorn temperature is by definition the temperature at which
the one-loop free energy (\ref{free}) diverges. This happens
for $\tau_2\to 0$. In this limit  it is useful to write the
Dedekind eta function in term of a series as in
\cite{Green:1987sp}
\be
\left|\eta(\frac{\tau_2}{2})\right|^{-24}=
e^{\pi \tau_2}
\left|\prod_{m=1}^\infty \left(1-e^{-\pi \tau_2
m}\right)\right|^{-24}
\label{etad}
\ee
and
\be
\prod_{d=1}^\infty \left(1-e^{-\pi \tau_2 m}\right)^{-24}\equiv
\sum_{r=0}^\infty d(r) e^{-\pi \tau_2r}
\label{prod} 
\ee
For large $r$ one gets~\cite{Huang:1970iq} 
\be
d(r)\sim r^{-27/4}e^{4\pi\sqrt{r}}
\label{asympt}
\ee
In the $\tau_2\to 0$ limit the sums are dominated by those
integers for which $r$ is big. Moreover, the dominant term is
obtained by setting $n=1$. Then for $\tau_2\sim 0$ we could use a
saddle point procedure for the variable $r$ to evaluate the sum
\begin{equation}
\sum_{r=0}^{\infty} r^{-27/4}
e^{4\pi\sqrt{r}-\pi\tau_2 r}
\label{saddle}
\end{equation}
The saddle point equation has the solution
\be
\sqrt{r}=\frac{2}{\tau_2}
\ee
Substituting the solution in
the expression of the free energy and taking the $\tau_2 \to 0$
limit, we find that the Hagedorn temperature is
\be
T=T_H\sqrt{(1-A^2)}
\label{hag}
\ee
where
$T_H=\frac{1}{4\pi\sqrt{\alpha'}}$ is the Hagedorn temperature in
the absence of the background metric.  
The result (\ref{hag}) is quite expected 
since it is exactly  the NCOS behavior
but with $B$ replaced by $-A$.

We now turn to examine the Hagedorn temperature when
both the metric (\ref{metric}) and
the constant Neveu-Schwarz $B$-field (\ref{bfield}) are present. 
The boundary conditions we choose are (\ref{bc}).

The string energy spectrum is
modified by the presence of the $B$-field and is given by
equation (\ref{spectrumb}).
In this case the expression for the free energy
becomes
\be
F=-\sum_{n=1}^{\infty}\int_0^{\infty}
\frac{d\tau_2}{\tau_2} \frac{\sqrt{1-A^2}}{(2\pi \alpha'\tau_2)^{\frac{D}{2}}}
\left|\eta\left(\frac{\tau_2}{2}\right)\right|^{-24}
e^{-\frac{\beta^2 n^2}{4\pi\alpha'\tau_2}\left(1-A^2 \right)
+\frac{\beta^2 n^2}{4\pi\alpha'\tau_2}B^2\left(1-A^2 \right)}
\label{bfree} 
\ee
We can then proceed as before and use a saddle point
procedure for the variable $r$. 
Substituting the solution of the saddle point equation in the
expression of the free energy one can easily see that the exponent
vanishes when 
\be
T=T_H\sqrt{(1-A^2)(1-B^2)}
\label{bhag}
\ee
where again $T_H$ is the Hagedorn temperature for the open bosonic
strings in the absence of the background metric and antisymmetric
tensor field. The Hagedorn temperature is 
self-dual when the duality transformation is along the spatial direction.
Whereas the Moyal phase depends only on one of the two background
moduli, the Hagedorn temperature depends on both.
By studying the high temperature behavior
of the theory with boundary conditions (\ref{bcab}) we obviously
find for the Hagedorn temperature the same result (\ref{bhag}),
the energy spectra are in fact dual.

The Hagedorn temperature can be determined also when the 
direction $1$ is compactified, it is easy to see that also in this case the 
result (\ref{bhag}) still holds and it coincides with the one 
obtained for the closed string sector~\cite{Grignani:2001ik}.
The formula (\ref{bhag}) in this case has a remarkable feature.  
It depends on $A$ and $B$,  
but for fixed $A$ and $B$, it does
not depend on the compactification radius $R$.  
The physical picture of what is happening has been given in
ref.~\cite{Grignani:2001ik}, here we review the argument.
When the boundary conditions are (\ref{bc}),
there is a region of the parameter space where $A$ and $B$ are between
0 and 1, away from their limiting values and where $R$ is very large
so that all wrapped states have a very large energy. In that case, at
temperatures just below $T_H$, practically no wrapped states are
excited in the thermal distribution.  However, since $T_H$ depends on
$B$, it must be wrapped states which condense at the Hagedorn
transition, in fact the resulting long string must wrap the compact
dimension.  
Thus we see that, in the limit where $R$ is very large, when the
temperature $T_H$ is reached, there is a catastrophic process where
dominant configurations in the ensemble go from a thermal distribution
of multi-string states with zero wrapping to a single long string
which wraps the compact dimension. The same considerations are valid
in the T-dual theory with boundary conditions (\ref{bcab}) provided 
the winding number is exchanged with the quantum of the momentum in the
compactified direction. 

In the thermal ensemble the total energy is proportional to the
volume, there is sufficient energy to produce strings of arbitrary length
whose energy only scale like their length.  Then, the
$R$-dependence of the total energy, which grows linearly in $R$ if the
temperature is held fixed, is similar to the energy
dependence of a wrapped string which also scales linearly with $R$.
There is always enough energy for a long string to wrap the compactified 
direction no matter how large is the compactification radius.

In~\cite{Klebanov:2000pp} it was noted
that, when the compactified dimension has finite radius,  
the wrapped closed string states do not
decouple in the NCOS limit.   These wrapped states get infinitely large
energy in the limit where the radius of the compact dimension is taken to 
infinity.  However, since the Hagedorn temperature does not depend on the
compactification radius, the
closed strings still participate in the Hagedorn transition. 
We see that, if the radius is very large but
finite, the phase transition for the NCOS, 
which is believed to be a second order phase transition,  
becomes first order for the presence of the closed strings 
in the spectrum.

\section{Conclusions}

In this Paper, we have studied the origin of space/time
noncommutativity in open string theory.
We have shown that when an open bosonic string propagates
in a background metric of the form (\ref{metric}) and with 
boundary conditions (\ref{bc}), a Moyal phase arises 
when calculating the OPE of vertex
operators.  The OPE needs a careful derivation of the propagators
of the theory and the correct definition of the vertex
operators.
The vertex operators are elements
of an algebra of function defined with
the Moyal $\ast$ product \cite{Seiberg:1999vs} 
instead of the usual commutative 
product.

In our approach the noncommutativity is not related to the presence 
of the antisymmetric tensor field $B_{\mu \nu}$, but it depends only 
on the metric moduli space.

We also study the effect of the background considered
on the Hagedorn temperature showing that it 
depends on both the parameters of the background metric and the 
antisymmetric tensor field.
The same dependence on the backgrounds is valid for the open and closed
string sector when the Dirichlet spatial direction is compactified. 

Recently there has been some attempts to study how noncommutativity arises
when more general backgrounds are considered.  
In~\cite{Dolan:2002px} a time dependent
noncommutativity parameter was computed in a model with 
a time-dependent background. In \cite{Mukhi:2002ck} it was shown
that starting with type IIB string theory on the pp-wave background 
with a compact lightlike direction and performing a T-duality
over the lightlike direction, one can go to a type IIA description
in terms of a non-relativistic closed string theory (NRCS)
which is tightly related to the model studied in this Paper.
It would be interesting to study these
more general cases in view of our results.

\section{Acknowledgments}

We have benefited from discussions with L. Griguolo, G. W. Semenoff.


\begin{thebibliography}{99} 


\bibitem{Witten:1985cc}
E.~Witten,
``Noncommutative Geometry And String Field Theory,''
Nucl.\ Phys.\ B {\bf 268}, 253 (1986).

\bibitem{Abouelsaood:gd}
A.~Abouelsaood, C.~G.~Callan, C.~R.~Nappi and S.~A.~Yost,
``Open Strings In Background Gauge Fields,''
Nucl.\ Phys.\ B {\bf 280}, 599 (1987).

\bibitem{Ambjorn:2000yr} J.~Ambjorn, Y.~M.~Makeenko, 
G.~W.~Semenoff and R.~J.~Szabo, ``String theory in electromagnetic 
fields,'' arXiv:hep-th/0012092. 

\bibitem{Douglas:2001ba}
M.~R.~Douglas and N.~A.~Nekrasov,
``Noncommutative field theory,''
Rev.\ Mod.\ Phys.\  {\bf 73}, 977 (2001)
[arXiv:hep-th/0106048].

\bibitem{Seiberg:1999vs} N.~Seiberg and E.~Witten, 
``String theory and noncommutative geometry,''
JHEP {\bf 9909}, 032 (1999) [arXiv:hep-th/9908142]. 

\bibitem{Seiberg:2000ms} N.~Seiberg, L.~Susskind and N.~Toumbas, 
``Strings in background electric field, space/time 
noncommutativity and a new noncritical string theory,'' JHEP {\bf 
0006}, 021 (2000) [hep-th/0005040]. 

\bibitem{Gopakumar:2000na} R.~Gopakumar, J.~Maldacena, S.~Minwalla 
and A.~Strominger, ``S-duality and noncommutative gauge theory,'' 
JHEP {\bf 0006}, 036 (2000) [hep-th/0005048]. 

\bibitem{Gomis:2000bn}
J.~Gomis, M.~Kleban, T.~Mehen, M.~Rangamani and S.~H.~Shenker,
``Noncommutative gauge dynamics from the string worldsheet,''
JHEP {\bf 0008}, 011 (2000)
[arXiv:hep-th/0003215].

\bibitem{Seiberg:2000gc}
N.~Seiberg, L.~Susskind and N.~Toumbas,
``Space/time non-commutativity and causality,''
JHEP {\bf 0006}, 044 (2000)
[arXiv:hep-th/0005015].

\bibitem{Barbon:2000sg}
J.~L.~Barbon and E.~Rabinovici,
``Stringy fuzziness as the custodian of time-space noncommutativity,''
Phys.\ Lett.\ B {\bf 486}, 202 (2000)
[arXiv:hep-th/0005073].

\bibitem{Alvarez-Gaume:2001ka}
L.~Alvarez-Gaume, J.~L.~Barbon and R.~Zwicky,
``Remarks on time-space noncommutative field theories,''
JHEP {\bf 0105}, 057 (2001)
[arXiv:hep-th/0103069].

\bibitem{Bassetto:2001vf}
A.~Bassetto, L.~Griguolo, G.~Nardelli and F.~Vian,
``On the unitarity of quantum gauge theories on noncommutative spaces,''
JHEP {\bf 0107}, 008 (2001)
[arXiv:hep-th/0105257].

\bibitem{Klebanov:2000pp} I.~R.~Klebanov and J.~Maldacena, ``1+1 
dimensional NCOS and its U(N) gauge theory dual,'' hep-th/0006085. 

\bibitem{Gomis:2000bd} J.~Gomis and H.~Ooguri, ``Non-relativistic 
closed string theory,'' hep-th/0009181. 

\bibitem{Danielsson:2000gi} U.~H.~Danielsson, A.~Guijosa and 
M.~Kruczenski, ``IIA/B, wound and wrapped,'' JHEP {\bf 0010}, 020 
(2000) [arXiv:hep-th/0009182]. 


\bibitem{Buscher:1988qj}T.~H.~Buscher, ``Path Integral Derivation 
Of Quantum Duality In Nonlinear Sigma Models,'' Phys.\ Lett.\ {\bf 
B201}, 466 (1988). 

\bibitem{Buscher:1987sk}T.~H.~Buscher, ``A Symmetry Of The String 
Background Field Equations,'' Phys.\ Lett.\ {\bf B194}, 59 (1987). 

\bibitem{Giveon:1994fu}
A.~Giveon, M.~Porrati and E.~Rabinovici,
``Target space duality in string theory,''
Phys.\ Rept.\  {\bf 244}, 77 (1994)
[arXiv:hep-th/9401139].

\bibitem{Grignani:2001ik} G.~Grignani, M.~Orselli and 
G.~W.~Semenoff, ``The target space dependence of the Hagedorn 
temperature,'' JHEP {\bf 0111}, 058 (2001) [arXiv:hep-th/0110152]. 

\bibitem{Chu:1998qz} C.~S.~Chu and P.~M.~Ho, ``Noncommutative open 
string and D-brane,'' Nucl.\ Phys.\ B {\bf 550}, 151 (1999). 

\bibitem{Horava:1989ga}
P.~Horava, ``Background Duality Of Open String Models,''
Phys.\ Lett.\ B {\bf 231}, 251 (1989).

\bibitem{Sakamoto:1989ig}
M.~Sakamoto,
``A Physical Interpretation Of Cocycle Factors In Vertex 
Operator Representations,''
Phys.\ Lett.\ B {\bf 231}, 258 (1989).

\bibitem{Green:1987sp} M.~B.~Green, J.~H.~Schwarz and E.~Witten, 
``Superstring Theory. Vol. 1: Introduction,'' {\it Cambridge, Uk: 
Univ. Pr. ( 1987) 469 P. ( Cambridge Monographs On Mathematical 
Physics)}. 


\bibitem{Polchinski:rq}
J.~Polchinski,
``String Theory. Vol. 1: An Introduction To The Bosonic String,''
{\it  Cambridge, UK: Univ. Pr. (1998) 402 p}.

\bibitem{Maharana:2000fc}
J.~Maharana and S.~S.~Pal,
Phys.\ Lett.\ B {\bf 488}, 410 (2000)
[arXiv:hep-th/0005113].

\bibitem{Grignani:1999sp} G.~Grignani and G.~W.~Semenoff, 
``Thermodynamic partition function of matrix superstrings,'' 
Nucl.\ Phys.\ {\bf B561}, 243 (1999) [hep-th/9903246]. 

\bibitem{Grignani:2000zm} G.~Grignani, P.~Orland, L.~D.~Paniak and 
G.~W.~Semenoff, ``Matrix theory interpretation of DLCQ string 
worldsheets,'' Phys.\ Rev.\ Lett.\ {\bf 85}, 3343 (2000) 
[hep-th/0004194]. 


\bibitem{Bowick:az}
M.~J.~Bowick and L.~C.~Wijewardhana,
Phys.\ Rev.\ Lett.\  {\bf 54}, 2485 (1985).

\bibitem{Alvarez:1986sj}
E.~Alvarez and M.~A.~Osorio,
Phys.\ Rev.\ D {\bf 36}, 1175 (1987).

\bibitem{Atick:1988si} J.~J.~Atick and E.~Witten, ``The Hagedorn 
Transition And The Number Of Degrees Of Freedom Of String 
Theory,'' Nucl.\ Phys.\ {\bf B310}, 291 (1988). 

\bibitem{Deo:1989bv} N.~Deo, S.~Jain and C.~Tan, ``String 
Statistical Mechanics Above Hagedorn Energy Density,'' Phys.\ 
Rev.\ D {\bf 40}, 2626 (1989). 

\bibitem{Ferrer:1990na}
E.~J.~Ferrer, E.~S.~Fradkin and V.~de la Incera,
``Effect Of A Background Electric Field Of The Hagedorn Temperature,''
Phys.\ Lett.\ B {\bf 248}, 281 (1990).

\bibitem{Tseytlin:1998kw}
A.~A.~Tseytlin,
``Open superstring partition function in constant gauge field background  
at finite temperature,''
Nucl.\ Phys.\ B {\bf 524}, 41 (1998)
[arXiv:hep-th/9802133].

\bibitem{Grignani:2001hb} G.~Grignani, M.~Orselli and 
G.~W.~Semenoff, ``Matrix strings in a B-field,'' hep-th/0104112. 


\bibitem{Gubser:2000mf} S.~S.~Gubser, S.~Gukov, I.~R.~Klebanov, 
M.~Rangamani and E.~Witten, ``The Hagedorn transition in 
non-commutative open string theory,'' hep-th/0009140. 

\bibitem{Barbon:2001tm} J.~L.~Barbon and E.~Rabinovici, ``On the 
nature of the Hagedorn transition in NCOS systems.'' JHEP {\bf 
0106}, 029 (2001) [hep-th/0104169]. 

\bibitem{Chan:2001gs} C.~S.~Chan, A.~Hashimoto and H.~Verlinde, 
``Duality cascade and oblique phases in non-commutative open 
string theory,'' hep-th/0107215.

\bibitem{Kogan:2001px}
I.~I.~Kogan, A.~Kovner and M.~Schvellinger,
``Hagedorn transition, vortices and D0 branes: 
Lessons from 2+1 confining  strings,'' JHEP {\bf 0107}, 019 (2001)
[arXiv:hep-th/0103235].

\bibitem{Huang:1970iq} K.~Huang and S.~Weinberg, ``Ultimate 
Temperature And The Early Universe,'' Phys.\ Rev.\ Lett.\ {\bf 
25}, 895 (1970). 


\bibitem{Dolan:2002px}
L. Dolan and C. R. Nappi,
``Noncommutativity in a time-dependent background,''
arXiv:hep-th/0210030.

\bibitem{Mukhi:2002ck}
S. Mukhi, M. Rangamani and E. Verlinde,
``Strings from Quivers, Membranes from Moose''
JHEP {\bf 0205}, 023 2002
[arXiv:hep-th/020204147]

\end{thebibliography}
\end{document}